# An evaporation-based model of thermal neutron induced ternary fission of plutonium

J. P. Lestone,
Applied Physics Division, Los Alamos National Laboratory,
Los Alamos, New Mexico 87545, USA
(Received May 16th 2006)

Ternary fission probabilities for thermal neutron induced fission of plutonium are analyzed within the framework of an evaporation-based model where the complexity of time-varying potentials, associated with the neck collapse, are included in a simplistic fashion. If the nuclear temperature at scission and the fission-neck-collapse time are assumed to be ~1.2 MeV and ~$10^{-22}$ s, respectively, then calculated relative probabilities of ternary-fission light-charged-particle emission follow the trends seen in the experimental data. The ability of this model to reproduce ternary fission probabilities spanning seven orders of magnitude for a wide range of light-particle charges and masses implies that ternary fission is caused by the coupling of an evaporation-like process with the rapid re-arrangement of the nuclear fluid following scission.

# I. Introduction

Ternary nuclear fission was first discovered in 1946 by Green and Livesey [1] and Tsien et al., [2] and is a process by which a third charged fragment is generated in the fission process close to the plane perpendicular to the direction of the two main charged fragments. The main ternary-fission observables include the probability per fission event of generating various isotopes, their energy spectra and angular distributions. Extensive measurements of these properties have been made and substantially documented in several review articles [3-6]. The dominant ternary-fission particles are $\alpha$ particles, which occur a few times per thousand thermal-neutron-induced or spontaneous fissions. The energy spectra and angular distributions of these particles show that they are generated close to the scission (rupture) point, between the two main fission fragments. Ternary fission is thus often referred to as near-scission emission. There is a very rare process that generates a third light charged fragment in the fission process close to the direction of motion of the main two fission fragments [5,6]. This is often called polar emission. To distinguish the dominant ternary-fission process from the rare polar emission, they are sometimes referred to as equatorial ternary fission (ETF) and polar ternary fission (PTF), respectively. In this paper we are mainly interested in equatorial ternary fission, and polar ternary fission is only briefly discussed.

Hot compound nuclei can evaporate neutrons and charged particles before fission, as can hot fission fragments after fission [7-9]. However, the energy cost for emitting ternary particles is so high that standard statistical evaporation can be ruled out [3,6]. This has led many authors to conclude that ternary fission is a dynamical process not associated with an evaporative process. Halpern [10] suggested that the sudden collapse of the neck material between the main two fragments produces a rapid change in the nuclear potential felt by light particles in the neck region. This rapid change in the nuclear potential can result in a large change of the potential energy of light particles in the neck region with little corresponding change in their kinetic energy [4]. However, the initial suggestion of Halpern has not been used to obtain a quantitative description of ternary fission, and many other dynamical models have been developed. These include models involving an extension of the theory of particle emission from actinide ground states to a rapidly evolving system in the last phase of the fission process [11], double-neck rupture [12], models where $\alpha$ particles inside the neck region gain energy from the average time-dependent potential of a fissioning system [13,14], and other dynamical models reviewed in ref. [6]. Although each of these dynamical models has had limited success in reproducing some of the features observed in the experimental data, no satisfactory simultaneous reproduction of a large amount of experimental data has been achieved.

More contemporary articles on ternary fission continue to support the view that ternary fission does not involve a standard statistical emission process. For example, the dependence of ternary fission on the charge of the light particle can be used to conclude that ternary fission is "inconsistent with a statistical emission mechanism in which emission barriers follow a standard $Z$ dependence" [15]. Recently, $\gamma$-$\gamma$-$\gamma$ coincidences and $\gamma$-$\gamma$-light charged particle coincidences have provided evidence for two different mechanisms for ternary fission, one hot and the other cold [16-18]. The $^{10}$Be accompanied fission of $^{252}$Cf appears to be dominated by the cold process [18].

A number of recent publications are consistent with ternary-fission emissions coming predominantly from a hot process. Isotope thermometry, which utilizes double isotope-yield ratios, has been used to infer apparent temperatures associated with ternary fission [19]. Yields of hydrogen, helium, lithium, and beryllium isotopes from the fission of heavy nuclei from $^{229}$Th to $^{252}$Cf, are used. The apparent temperature for low-energy ternary fission is 1.10±0.15 MeV. The yield of excited $^{8}$Li ions [20] and the yield of 3.368 MeV gamma rays from the first excited state of $^{10}$Be [21] in the ternary fission of $^{252}$Cf suggest that these ions are being generated by a hot process with a temperature of ~1 MeV. The ratios of yields of ternary light-charged particles (LCP) from fissioning systems differing by two neutrons suggest that low-energy ternary fission involves a statistical process where the ejected particles are in equilibrium with a heat bath with a temperature slightly hotter than 1 MeV [22]. A scission configuration with a temperature of ~1 MeV is consistent with the surface-plus-window dissipation model [23,24] for the conversion of nuclear collective energy into internal degrees of freedom. The surface-plus-window dissipation mechanism is in agreement with experimental mean fission-fragment kinetic energies for a wide range of fissioning nuclei, and the widths of isoscalar giant quadrupole and giant octupole resonances. More recent calculations have confirmed that surface-plus-window dissipation is consistent with the experimental mean kinetic energy of fission fragments [25]. Some authors believe that the asymmetric mass distribution and the preference for even-proton fission fragments in low-energy fission proves that the temperature in low-energy fission must be significantly less than 1 MeV because microscopic and pairing corrections are very important, and have therefore dismissed the possibility of the hot scission configuration. The asymmetric mass distribution and the preference for even-proton fragments only proves that the temperature must be low at some point along the path to fission, but does not rule out the possibility that collective flow is converted into significant heat during the final stages of the fission process, just before scission (but after the mass and proton-number distributions have been set), as predicted by the surface-plus-window dissipation model.

Based on the high energy cost to produce ternary fission, it is clear that this process cannot be associated with standard particle evaporation from hot nuclear matter. Many have used this fact to rule out an evaporative process, instead of considering how the nature of an evaporative process might change in the presence of a rapidly moving hot nuclear fluid. A combined statistical and dynamical model of ternary fission was recently introduced [24]. In this model, statistical theory is used to calculate the probability that particles are evaporated from the nuclear surface with insufficient energy to surmount the Coulomb barrier. These quasi-evaporated particles exist between the nuclear surface and the Coulomb barrier for a short period of time before returning to the nuclear fluid. Potential ternary-fission particles are first quasi-evaporated into the region surrounding the pre-scission neck material. Then, due to the rapid collapse of the neck material, quasi-evaporated particles above the neck-rupture location experience a rapid rise in their nuclear potential and are ejected perpendicular to the direction of the main fragments via a purely classical process. This particle emission mechanism can be viewed as a coupling of the sudden approximation first suggested by Halpern [10] and particle evaporation, and has been used to explain many of the properties of $^{235}$U($n_{th}$,$f$) ternary fission [24]. However, others have suggested that the classical concepts used in the combined statistical and dynamical model are invalid, and that the results shown in Figs 3-5 of ref. [24] are not evidence that an evaporative process is occurring in low-energy ternary fission. It appears that the suggestion that ternary fission is associated with an evaporation process is not being taken seriously because of concerns related to the concept of quasi-evaporated particles introduced in ref [24]. In the present paper, a more simplistic approach is used to couple the standard theory of particle evaporation with a rapid neck collapse. It is hoped that this will provide transparent evidence that a process with evaporation-like properties is playing a central role in low-energy ternary fission.

## II. The model

In the statistical limit, the particle evaporation rate per unit area from non-accelerating hot compound nuclei is [26]

$$R = \frac{(2s+1)\mu T^2}{4\pi^2\hbar^3} \exp\left(-\Delta E / T\right),  \tag{1}$$

where $s$ is the spin of the evaporated particle, $\mu$ is the reduced mass of the particle-daughter system, and $\Delta E$ is the energy cost of getting a particle to the emission barrier. $\Delta E$ is the particle binding energy plus the potential energy of the LCP at the corresponding emission barrier. Based on the terms in Eq. (1), we postulate that, if ternary fission is associated with an evaporation-like process, the LCP emission probability from a given fissioning system, with fixed initial excitation energy, is of the form

$$P_{TF} = C\,(2s+1)\,A_{LCP}\,\exp\!\left(\frac{-\Delta E}{T_{\mathrm{scis}}}\right), \tag{2}$$

where $C$ is a constant, and $T_{\mathrm{scis}}$ is the temperature at scission. One must be careful when defining the particle binding energies for ternary fission and when calculating the potential energy at the corresponding emission barriers.

## II.A Particle binding energies

The particle binding energy

$$B_E = M_D + M_{LCP} - M_P, \tag{3}$$

where the subscripts "$P$" and "$D$" refer to the parent and daughter systems, is well defined if the parent and daughter are ground-state nuclei. However, the situation becomes more complex for emission from hot deformed systems because the shapes of both the parent and daughter need to be defined. If the parent and daughter are sufficiently hot then shell and pairing corrections can be neglected and the non-thermal mass of the parent and daughter can be estimated via the liquid drop model and expressed as the sum of volume, surface, and Coulomb terms,

$$M_{LDM} = E_V + E_S + E_C. \tag{4}$$

The volume, surface, and Coulomb terms can be expressed as [27]

$$E_V = -15.4941\ \mathrm{MeV}\,(1 - 1.7828(\frac{A - 2Z}{A})^2)\,A, \tag{5}$$

$$E_S = B_S\ 17.9439\ \mathrm{MeV}\,(1 - 1.7828(\frac{A - 2Z}{A})^2)\,A^{2/3},\ \mathrm{and} \tag{6}$$

$$E_C = B_C\ 0.7053\ \mathrm{MeV}\,\frac{Z^2}{A^{1/3}} - 1.1529\ \mathrm{MeV}\,\frac{Z^2}{A}, \tag{7}$$

where, $B_S$ and $B_C$ are surface and Coulomb scaling factors that depend on the assumed nuclear shape. For spherical nuclei, both of these scaling factors are 1.0. For deformed shapes, $B_S$ is greater than one, and $B_C$ is less than one. The surface-plus-window dissipation model [23] predicts that scission occurs at a distance between mass centers of the nascent fragments $d_{\mathrm{scis}} \sim 2.6$ times the nuclear radius of the corresponding spherical system ($R_o$), assuming symmetric fission. Here, we assume the same distance between mass centers at scission. The surface and Coulomb scaling factors for this assumed scission configuration can be estimated using the Modified Liquid Drop Model, MLDM [28]. The corresponding values are $B_S$=1.272 and $B_C$=0.784. The surface and Coulomb energies of spherical systems are many hundreds of MeV and, therefore, the surface and Coulomb shape dependent scaling factors must be determined to at least the third decimal point to accurately define the potential energy surface of a single system. Here, we are only interested in calculating particle binding energies. The particle binding energies are determined using the difference between the parent and daughter nuclear potential energies and, therefore, even several percent inaccuracies in $B_S$ and $B_C$ do not significantly change the nuclear temperature at scission and the neck collapse time determined in section III of this paper. Later in this section particle binding energies are calculated, assuming that scission configurations can be represented by two spherical liquid drop model (LDM) fission fragments. Even using this gross approximation does not significantly affect the conclusions drawn from the present work.

An initial estimate of the particle binding energies can be obtained using Eqs. (3) through (7) with the values for surface and Coulomb scaling factors appropriate to the scission configuration given above. These estimates of the particle binding energies assume that the parent and daughter systems have the same distance between mass centers in units of the corresponding spherical systems. We shall refer to these estimates as the particle binding energies assuming a fixed shape for both the parent and daughter systems. However, the deformation of the daughter following emission from the scission configuration will depend on the location from which the LCP is evaporated. To illustrate this effect, consider the emission of an α particle from the $^{242}$Pu scission configuration. The distance between mass centers of the parent system is 19.85 fm, assuming the radius of the corresponding spherical system is 1.225 fm × $242^{1/3} = 7.63$ fm. The radius parameter of 1.225 fm was chosen because it is consistent with the assumed Coulomb energy given by Eq. (7). The corresponding distance between mass centers of the same shaped $^{238}$U daughter is $(238/242)^{1/3} \times 19.84 = 19.73$ fm. For equatorial ternary fission, the parent scission configuration generates LCP in the region between the nascent fragments. If the $^{242}$Pu scission configuration generates an α particle close to the plane between the two nascent fragments then the distance between the $^{238}$U nascent (daughter) fragments must be larger than the separation of the $^{242}$Pu nascent fragments. For α particles generated close to the plane between the two nascent fragments, the distance between mass centers of the $^{238}$U daughter nascent fragments is ~19.85 fm × 242/238 = 20.18 fm. This increase in the separation of the $^{238}$U daughter nascent fragments relative to the value of 19.73 fm obtained assuming a fixed shape for both the parent and daughter causes a decrease (shift) in the particle binding energies for equatorial ternary fission. This shift can be expressed as follows [22]:

$$B_E^{Shift} = 0.36 \, (\text{MeV} \cdot \text{fm}) \, \frac{Z_D^2}{d_{scis}} \left\{ \frac{A_D}{A_P} - \left( \frac{A_P}{A_D} \right)^{1/3} \right\}, \tag{8}$$

where $d_{scis}$ is the distance between the mass centers of the nascent fragments of the parent, and $Z_D$, $A_D$, and $A_P$ are the atomic and mass numbers for the daughter and parent systems. Given that the mass of the fissioning systems are many times larger than the mass of the LCP, Eq. (8) is well approximated by

$$B_E^{Shift} = -0.15 \, (\text{MeV}) \, \frac{A_{LCP} Z_D^2}{A_P^{4/3}}. \tag{9}$$

The proportionality constant in Eq. (9) scales with the assumed distance between mass centers of the nascent fragments at the moment of scission. This is presently defined by the product of the assumed radius parameter (1.225 fm) and the assumed distance between mass centers of the nascent fragments at the moment of scission in units of the radius of the corresponding spherical system (2.6). The distance between mass centers of the nascent fragments at the moment of scission is not a precisely known quantity and therefore the proportionality constant in Eq. (9) has an uncertainty of ~10-20%. An adjustment to the proportionality constant in Eq. (9) is considered in section III. Table I contains particle binding energies for various $Z$=1 to 12 particles generated between the nascent fragments of a $^{242}$Pu system at the moment of scission.

TABLE I. Various particle binding energies for the scission configuration of $^{242}$Pu assuming the same shape for parent and daughter systems (fixed shape); the shift in these binding energies (as discussed in the text); and the corresponding binding energies for Equatorial Ternary Fission (ETF).

| Isotope | $B_E^{Scission}$ fixed shape (MeV) | $B_E^{Shift}$ ETF (MeV) | $B_E$ ETF (MeV) | Isotope | $B_E^{Scission}$ fixed shape (MeV) | $B_E^{Shift}$ ETF (MeV) | $B_E$ ETF (MeV) |
|---|---|---|---|---|---|---|---|
| $^1$H | 8.72 | -0.86 | 7.85 | $^{14}$C | -8.63 | -10.84 | -19.46 |
| $^2$H | 11.85 | -1.73 | 10.12 | $^{20}$C | 10.72 | -15.48 | -4.76 |
| $^3$H | 11.07 | -2.59 | 8.47 | $^{16}$N | -6.64 | -12.10 | -18.74 |
| $^4$He | -0.02 | -3.38 | -3.40 | $^{21}$N | 0.84 | -15.84 | -15.05 |
| $^8$He | 19.15 | -6.77 | 12.39 | $^{19}$O | -12.43 | -14.04 | -26.48 |
| $^7$Li | 8.68 | -5.79 | 2.89 | $^{22}$O | -14.29 | -16.26 | -30.55 |
| $^{11}$Li | 24.86 | -9.10 | 15.75 | $^{19}$F | -11.67 | -13.72 | -25.39 |
| $^9$Be | 4.14 | -7.29 | -3.15 | $^{22}$F | -16.29 | -15.29 | -32.18 |
| $^{14}$Be | 21.46 | -11.33 | 10.12 | $^{24}$Ne | -25.51 | -16.92 | -42.43 |
| $^{11}$B | 0.61 | -8.71 | -8.10 | $^{28}$Na | -26.69 | -19.28 | -45.97 |
| $^{15}$B | 10.48 | -11.87 | -1.40 | $^{30}$Mg | -35.61 | -20.16 | -55.77 |

To test the sensitivity of the model calculations presented here to the details of how the particle binding energies are calculated, the particle binding energies for equatorial ternary fission have been estimated by two additional methods. In the first of these alternative methods the parent scission configuration is assumed to be well represented by two identical spherical LDM nuclei with a distance between mass centers equal to 2.6 times the radius of the fissioning compound system. The mass of the parent scission configuration is then given by

$$M_P = 2 \times M_{LDM}(Z_P/2, A_P/2) + \frac{0.113 \,(\text{MeV})Z_P^2}{A_P^{1/3}} . \qquad (10)$$

The mass of the corresponding daughter can be estimated by

$$M_D = 2 \times M_{LDM}(Z_D/2, A_D/2) + \frac{0.113 \,(\text{MeV})Z_D^2 A_D}{A_P^{4/3}} . \qquad (11)$$

The distance between mass centers of the daughter fragments is larger than the corresponding distance for the parent fragments, by a factor of $A_P/A_D$, because the nucleons needed to form the equatorial LCP are assumed to move from the parent fragments to a location symmetrically between the main fragments. This assignment of the LCP nucleons to the region between the main fragments causes an increase in the distance between mass centers of the daughter fragments relative to the parent fragments.

In the second alternative method any sensitivity of the conclusions drawn from the present work on the assumption of mass symmetric fission is tested by estimating the particle binding energies for equatorial ternary fission assuming non-identical spherical LDM nuclei with a distance between mass centers equal to 2.6 times the radius of the fissioning compound system. The mass of the parent system is then given by

$$M_P = M_{LDM}(Z_H, A_H) + M_{LDM}(Z_L, A_L) + \frac{0.452 \,(\text{MeV})Z_H Z_L}{A_P^{1/3}} . \qquad (12)$$

The mass and atomic numbers of the parent heavy and light scission fragments are assumed to be $A_H$=140, $Z_H$ =140 $\times Z_P/A_P$, $A_L= A_P - A_H$, and $Z_L = Z_P - Z_H$. If the mass of the LCP is assumed to come equally from both parent fragments and the LCP is assumed to be formed symmetrically between the main fragments then the mass of the corresponding daughter can be estimated by

$$M_D = M_{LDM}\left(Z_H - \frac{Z_{LCP}}{2}, A_H - \frac{A_{LCP}}{2}\right) + M_{LDM}\left(Z_L - \frac{Z_{LCP}}{2}, A_L - \frac{A_{LCP}}{2}\right)$$

$$+ \frac{0.452 \text{ (MeV)}\left(Z_H - \frac{Z_{LCP}}{2}\right)\left(Z_L - \frac{Z_{LCP}}{2}\right)}{A_P^{1/3}\left(\frac{A_H}{2A_H - A_{LCP}} + \frac{A_L}{2A_L - A_{LCP}}\right)} \qquad . \ (13)$$

Other reasonable assumptions about where the LCP is formed and what fraction of the LCP comes from which of the parent fragments do not significantly affect the result. Table II contains a comparison of the particle binding energies for various $Z=1$ to 12 particles generated between the nascent fragments of a $^{242}$Pu system at the moment of scission, estimated using the three approaches discussed above. All the calculations of the relative ternary fission emission probabilities presented in section III use the particle binding energy estimates obtained assuming MLDM scission configurations (see table I, or model I in table II). The other two methods described above (see models II and III in table II) produce negligible changes in the quality of the agreement between the model calculations and the experimental data, and negligible changes in the extracted nuclear temperature at scission and the neck collapse time (see section III).

TABLE II. A comparison of equatorial ternary fission particle binding energies for the scission configuration of $^{242}$Pu obtained using MLDM scission configurations as presented in table I (model I); assuming two identical spherical fragments to represent scission (model II); and assuming two non-identical spherical fragments to represent scission with the mass number of the heavier of the parent fragments equal to 140 (model III).

| LCP | Model I (MeV) | Model II (MeV) | Model III (MeV) | LCP | Model I (MeV) | Model II (MeV) | Model III (MeV) |
|---|---|---|---|---|---|---|---|
| $^1$H | 7.85 | 7.79 | 7.79 | $^{14}$C | -19.46 | -19.89 | -20.40 |
| $^2$H | 10.12 | 10.06 | 10.05 | $^{20}$C | -4.76 | -5.26 | -6.35 |
| $^3$H | 8.47 | 8.42 | 8.39 | $^{16}$N | -18.74 | -19.25 | -19.91 |
| $^4$He | -3.40 | -3.53 | -3.57 | $^{21}$N | -15.05 | -15.62 | -16.81 |
| $^8$He | 12.39 | 12.27 | 12.09 | $^{19}$O | -26.48 | -27.08 | -28.02 |
| $^7$Li | 2.89 | 2.70 | 2.56 | $^{22}$O | -30.55 | -31.20 | -32.49 |
| $^{11}$Li | 15.75 | 15.55 | 15.22 | $^{19}$F | -25.39 | -26.05 | -26.94 |
| $^9$Be | -3.15 | -3.41 | -3.62 | $^{22}$F | -32.18 | -32.89 | -34.14 |
| $^{14}$Be | 10.12 | 9.83 | 9.29 | $^{24}$Ne | -42.43 | -43.24 | -44.70 |
| $^{11}$B | -8.10 | -8.44 | -8.76 | $^{28}$Na | -45.97 | -46.91 | -48.91 |
| $^{15}$B | -1.40 | -1.76 | -2.38 | $^{30}$Mg | -55.77 | -56.81 | -59.08 |

## II.B Emission barrier heights

To be classically evaporated from a non-accelerating nuclear fluid, a charged particle must have enough energy to reach a point above the nuclear surface where the Coulomb repulsion and the nuclear attraction are equal. The barrier location for the evaporation of LCP from a spherical non-evolving hot compound system is easily defined. First, the potential energy of the LCP in the field of the daughter system is calculated as a function of the distance from the center-of-mass of the daughter. The emission barrier is located several Fermi beyond the nuclear surface where the derivative of the potential with respect to the distance from the center-of-mass is zero. This position is ~$r_p+3\delta$ [29] from the surface of nearly spherical nuclei where $r_p$ is the radius of the particle and $\delta$ is the diffuseness of the Woods-Saxon nuclear potential. A classical point particle placed at rest beyond the barrier location will be accelerated away from the daughter (ejected), while a particle placed at rest inside the barrier location will be accelerated towards the daughter system (absorbed). Fig. 1 shows an estimate of the $^{242}$Pu scission configuration (assuming symmetric fission) and the location of the fission fragments immediately following scission assuming an instantaneous collapse to spherical fragments. At the time of scission the potential energy of LCP near the plane perpendicular to and between the two main nascent fragments will be governed by both nuclear and Coulomb forces. As the neck material collapses, the influence of the nuclear force between the two main fragments will diminish with time. On the plane perpendicular to and between the two main nascent fragments, the Coulomb potential will dominate after the neck collapse is completed.

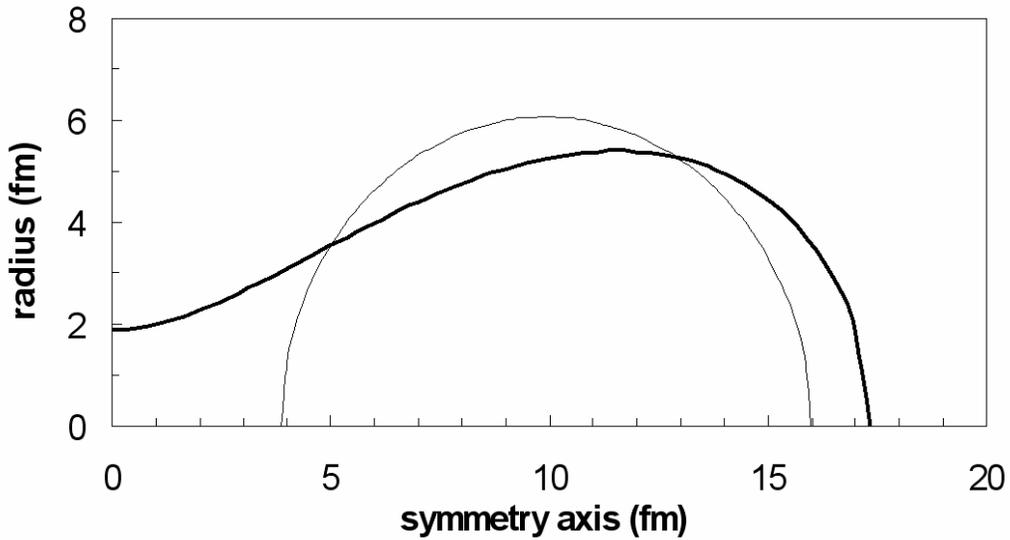

Fig. 1. The thick solid curve represents a $^{242}$Pu scission configuration. The assumed spherical fragments, immediately following scission, are represented by the thin solid curve.

Fig. 2 shows an estimate of the potential energy of an $\alpha$ particle in the plane perpendicular to and between two nascent symmetric fission fragments at the moment of scission for $^{242}$Pu. The Coulomb potential energy is estimated using the expression

$$C_P(r) = 1.44 \, (\text{MeV} \cdot \text{fm}) \frac{Z_D \, Z_{LCP}}{\sqrt{r_{scis}^2 + r^2}}, \tag{14}$$

where $r$ is the distance from the symmetry axis and $r_{scis}$ is $d_{scis}/2$. The nuclear potential is assumed to be

$$N_P(r) = \frac{-V_o}{1 + \exp(\frac{r - r_o}{\delta})}, \tag{15}$$

with a depth $V_o$=50 MeV, and $\delta$=0.6 fm [30]. The radius parameter of $r_o$=4 fm was chosen here because this is approximately the sum of the expected neck radius at scission and the radius of an $\alpha$ particle. If the potential shown in Fig. 2 is assumed static then the barrier location would be $r_B$=6.6 fm, with a barrier height $V_B$=21.6 MeV. The evolution of $\alpha$ particles starting at rest, at various values of $r$, is shown in Fig. 3. Notice that $\alpha$ particles with starting locations with $r < r_B$ fall inwards, while those with starting locations beyond $r_B$ are ejected. This behavior is obvious in the case of a static barrier. If the radius of the nuclear potential around the scission location is assumed to be ~4 fm for all LCP then the corresponding static barrier heights are ~11 MeV × $Z_{LCP}$. However, if the nuclear fluid is accelerating relative to a particle attempting to escape, then an effective barrier location can be defined that can be significantly different from the static barrier location. Plots similar to Fig. 3 can be used to define an effective barrier location for a non-static system.

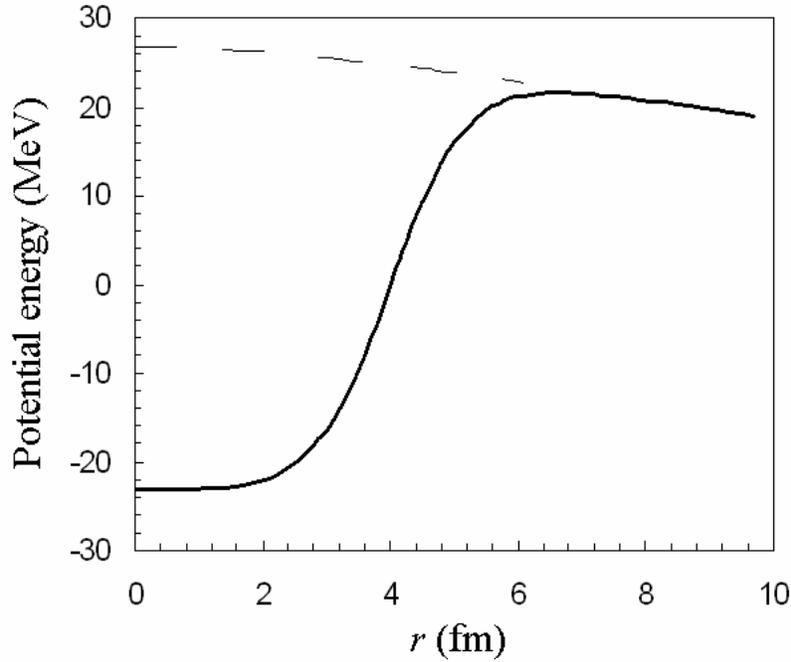

Fig. 2. The potential energy of an α particle in the plane perpendicular to and between two symmetric fission fragments at the moment of scission of $^{242}$Pu (solid curve). The dashed curve shows the Coulomb potential inside the emission barrier.

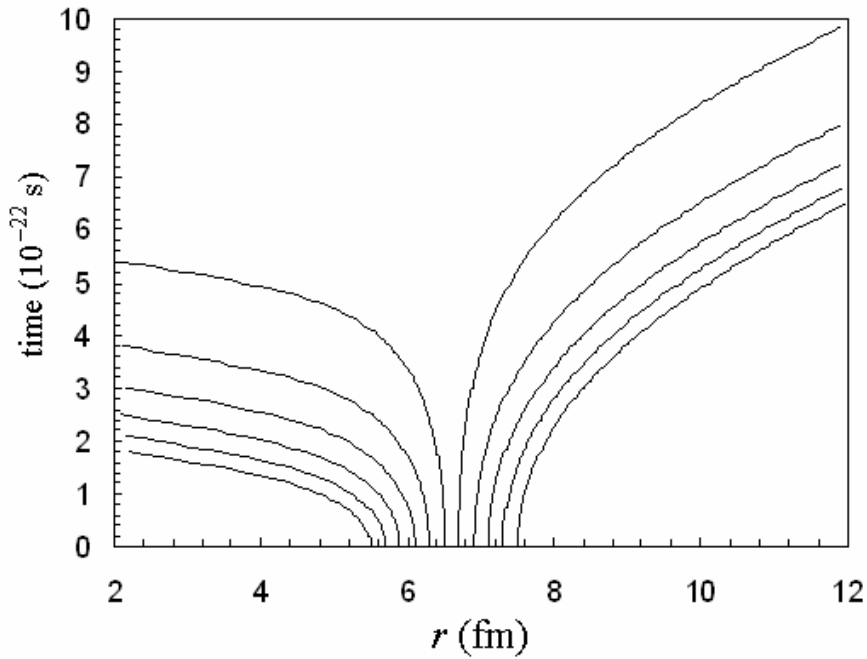

Fig. 3. The evolution of α particles (starting at rest) in the plane between the nascent fragments and perpendicular to the symmetry axis for a static $^{242}$Pu scission configuration, for various starting locations ($r$).

Immediately following scission, the potential energy of a particle between the nascent fragments will change rapidly with time as the neck collapses. To model this rapid potential change in a simplistic fashion, we shall initially assume that the radius parameter of the Woods-Saxon nuclear potential around the neck varies linearly from 4 fm at $t=0$, to zero at a time $t_{NC}$ (neck collapse time). This choice for the time dependence of the nuclear potential is very crude and leads to a cusp in the nuclear potential at $r=0$ as the time approaches $t_{NC}$. However, this artificial property of the chosen time dependent nuclear potential is not used in determining the effective barrier locations. The effective barriers are defined by the impulse that the LCP receives from the nuclear potential at radii larger than 3 fm and the Coulomb repulsion at times longer than the neck collapse time. The neck collapse time used here defines an effective velocity of the radius of the nuclear potential in the neck region soon after the neck breaks (scission). For example, with $t_{NC}=10^{-22}$ s the radius parameter of the nuclear potential between the nascent fragments, following scission, has a velocity of $4\times10^{22}$ fm/s. This is approximately half the Fermi velocity of nucleons inside nuclear matter. The evolution of $\alpha$ particles starting at rest, for various starting values of $r$, with an assumed neck collapse time, $t_{NC}=10^{-22}$ s, are shown is Fig. 4. For all starting locations with $r$ inside the static barrier location, $r_B$, the initial motion is inwards. However, as the neck collapses, the static barrier location moves inwards and some of the particles that start inside the static barrier location can find themselves outside the static barrier location of the evolving post-scission system. For the case illustrated in Fig. 4, $\alpha$ particles with starting locations $r < 4.5$ fm continue to fall inwards while $\alpha$ particles with starting locations with $r > 4.5$ fm are ejected in an outwards direction. The effective barrier location is, therefore, $r_{EB}=4.5$ fm. The corresponding effective barrier height is $V_{EB}=9$ MeV. The Coulomb and nuclear components of this effective barrier height are 24 MeV and –15 MeV, respectively. In the ternary-fission model presented here, it is assumed that this dramatic decrease in the effective emission barrier (between the nascent fragments) at the time of scission is the cause of equatorial ternary fission.

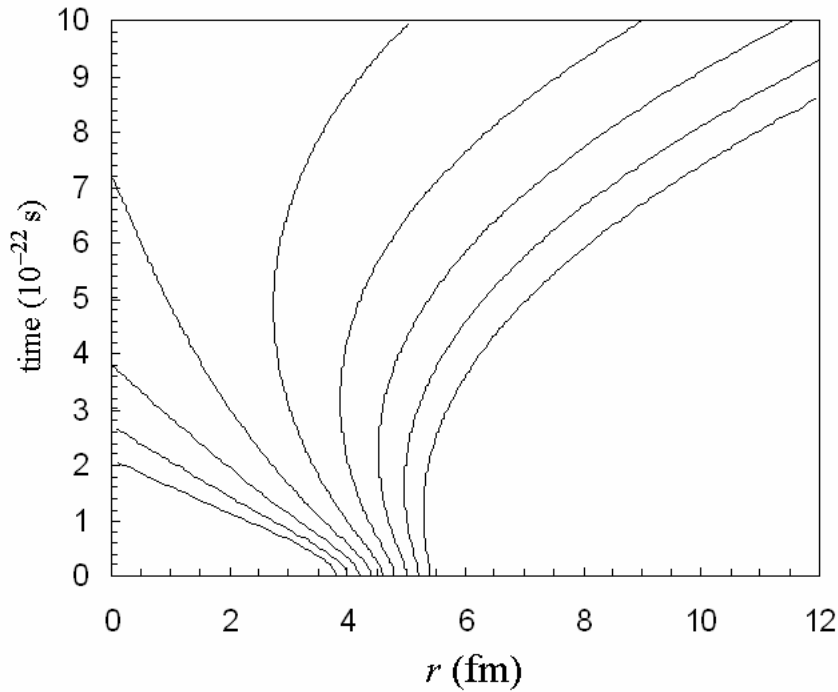

Fig. 4. The evolution of $\alpha$ particles (starting at rest) in the plane between the nascent fragments and perpendicular to the symmetry axis following the scission of $^{242}$Pu, assuming the moving Woods-Saxon nuclear potential discussed in the text. The neck collapse time is assumed to be $t_{NC}=10^{-22}$ s.

Using Eqs. (14) and (15) and the assumed linear variation of the radius parameter of the Woods-Saxon nuclear potential (discussed above) over a neck collapse time $t_{NC}$, the effective equatorial ternary fission barrier location, and thus the potential energy at the effective barrier location, can be determined by numerical means as a function of both $Z_{LCP}$ and $A_{LCP}$. Fig. 5 shows the decrease in the effective barrier heights relative to the standard static barrier heights for equatorial ternary fission of $^{242}$Pu as a function of neck collapse time for various LCP. For neck collapse times much longer than $10^{-21}$ s the effective barrier heights for ternary fission are not significantly affected by the neck collapse. However, the drop in effective barrier heights relative to the static barriers at scission becomes significant as the neck collapse time decreases below $10^{-21}$ seconds. This lowering of the effective barrier heights is a function of both the $A$ and $Z$ of the LCP, and for the lightest particles ($Z<7$) is approximately proportional to the square-root of the product of the light particle charge and mass. The dependence of $\Delta V$ on the square root of the product of $Z_{LCP}$ and $A_{LCP}$ is caused by a competition between the nuclear and Coulomb forces. If the neck collapse is rapid, then the motion of LCP that start at rest inside the static barrier at scission is governed by the nuclear force immediately following scission and by Coulomb repulsion at longer times. The inward kinetic energy gained through the action of the nuclear force is approximately proportional to the square of the initial depth below the static barrier divided by the mass of the LCP. If the Coulomb repulsion at the later times can stop the LCP from reaching the symmetry axis then the initial starting location is assumed to be beyond the effective barrier location (see Fig. 4). The product of $Z_D$ and $Z_{LCP}$ governs the Coulomb repulsion.

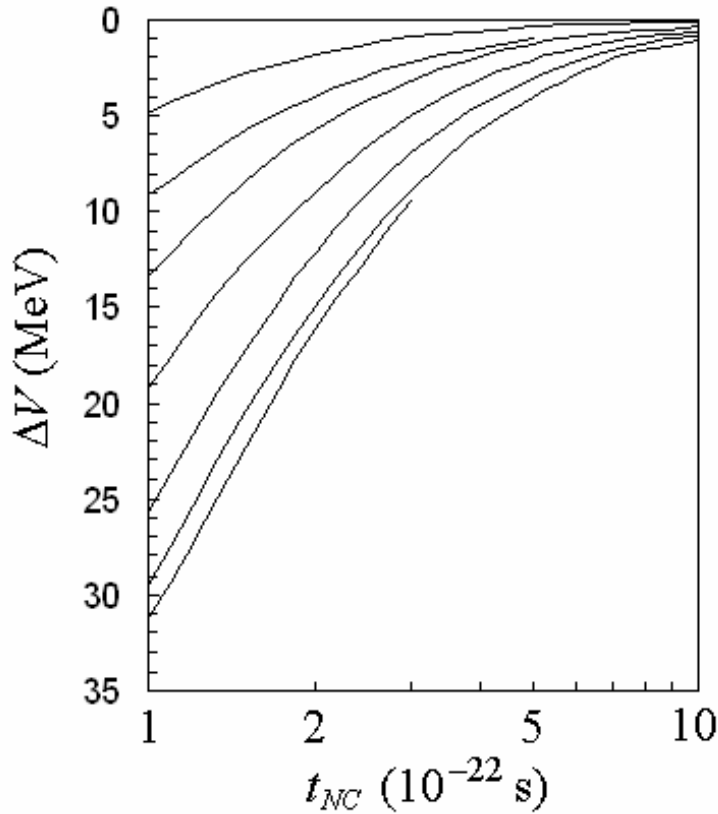

Fig. 5. The decrease in the effective barrier heights relative to the standard static barrier heights for equatorial ternary fission from $^{242}$Pu as a function of neck collapse time for various LCP. The curves, from upper to lower, are for protons, tritons, $\alpha$ particles, $^{8}$He, $^{10}$Be, $^{14}$Be and $^{14}$C, respectively.

By further simplification of the model, it is possible to obtain an analytical expression for relative ternary fission probabilities as a function of a temperature, a time scale, and assumed properties of the potential of LCP in the neck region. In order to obtain an analytical expression we assume that in the neck region, at the moment of scission, the nuclear force is zero beyond a barrier radius of $r_B$ and a constant value of $F_N$ for all locations inside the barrier radius. We assume that the barrier radius varies linearly from $r_B$ at scission to a value of zero at a time $t_{NC}$ following scission. Given these assumptions, the standard static barriers at scission can be obtained using Eq. (14) with $r = r_B$. Given the above-discussed simplifying assumptions, the effective emission barrier heights are lower than the static barriers by an amount (see appendix A)

$$\Delta V = \frac{2 F_N r_B E_B}{\Delta V_o} \left( \frac{\Delta V_o}{2 E_B} - \left( 1 + \frac{F_N r_B}{\Delta V_o} \right) + \sqrt{\left( 1 + \frac{F_N r_B}{\Delta V_o} \right)^2 - \frac{F_N r_B}{E_B}} \right), \tag{16}$$

where $E_B$ is the kinetic energy of the LCP if it was moving with the velocity of the barrier ($r_B \div t_{NC}$), and $\Delta V_o$ is the decrease in the barrier heights in the limits $t_{NC} \to 0$ and $F_N \to \infty$, and is given by

$$\Delta V_o = 0.61 \frac{\text{MeV } 10^{-22} \text{s}}{\text{fm}^2} \frac{r_B^2}{t_{NC}} \sqrt{\frac{Z_D Z_{LCP} A_{LCP}}{A_P}}. \tag{17}$$

Even though Eq. (16) was derived using the simplifying assumptions described above, it can be used to mimic the behavior of more complex potentials. For example, with $r_B = 5.2$ fm, Eq. (16) with $t_{NC} = 1.43 \times 10^{-22}$ s and $2.95 \times 10^{-22}$ s, mimics the $1 \times 10^{-22}$ s and $2 \times 10^{-22}$ s results shown in Fig. 5, with $F_N = 13.8$ MeV/fm and 10.1 MeV/fm, respectively. The times required to mimic the sliding Woods-Saxon potential (see Fig. 5) with Eq. (16) are ~50% longer than the true values. By numerical simulation it can be shown that other assumed spatial and temporal dependencies of the nuclear potential in the neck region, at and following scission, can be mimicked by Eq. (16). If the nuclear potential around the neck, at and following scission, is assumed to be of the form of an exponential shifting linearly with time:

$$V_N(r,t) = -V_o \exp \left( \frac{r_o (1 - t/t_{NC}) - r}{\delta} \right), \tag{18}$$

then the results for $\Delta V$ obtained numerically with neck collapse times of the order of $10^{-22}$ s, $r_o = 4$ fm, $\delta = 0.6$ fm, and $V_o = 50$ MeV $\times A_{LCP}$, can be mimicked using Eq. (16) with $r_B = 5.2$ fm and collapse times ~25% longer than the true values. Assuming the spatial and temporal dependencies of the nuclear potential as given in Eq. (18) it can be shown in the limits as $t_{NC} \to 0$ and $V_o \to \infty$ that (see Appendix A)

$$\Delta V_o \sim 0.61 \frac{\text{MeV } 10^{-22} \text{s}}{\text{fm}^2} \frac{r_o \, r_B}{t_{NC}} \sqrt{\frac{Z_D Z_{LCP} A_{LCP}}{A_P}}. \tag{19}$$

Notice the similarities between Eqs. (19) and (17). These two equations differ by $r_B/r_o$ and thus the ~25% increase in the times needed for Eq. (16) to mimic the $\Delta V$ associated with Eq. (18).

If the nuclear potential around the neck, at and following scission, is assumed to be of the form of an exponential weakening linearly with time:

$$V_N(r,t) = -V_o \left( \frac{t_{NC} - t}{t_{NC}} \right) \exp \left( \frac{r_o - r}{\delta} \right), \ (t \le t_{NC}), \tag{20}$$

then the results for $\Delta V$ obtained numerically with neck collapse times of the order of $10^{-22}$ s , $r_o = 4$ fm, $\delta = 0.6$ fm, and $V_o = 50$ MeV $\times A_{LCP}$, can be mimicked using Eq. (16) with $t_{NC}$ ~4 times the true values. Assuming the nuclear potential as given in Eq. (20), it can be shown in the limits as $t_{NC} \to 0$ and $V_o \to \infty$ that (see Appendix A)

$$\Delta V_o \sim 1.22 \frac{\text{MeV } 10^{-22}\text{s}}{\text{fm}^2} \frac{\delta r_B}{t_{NC}} \sqrt{\frac{Z_D Z_{LCP} A_{LCP}}{A_P}} . \tag{21}$$

Eq. (21) differs from Eq. (17) by $r_B/(2\delta)$, and thus the time scaling factor of ~4 needed for Eq. (16) to mimic the $\Delta V$ associated with the potential given by Eq. (20). Similar expressions can be derived assuming an exponential nuclear potential accelerating uniformly through space as a function of time, or weakening quadratically with time, or weakly exponentially with time. In fact, for any reasonable spatial and temporal dependence of the nuclear potential in the region of the neck at and following scission, the decrease in the barrier heights, in the limits as $t_{NC} \to 0$ and the nuclear force inside the barrier$\to \infty$, can be expressed as

$$\Delta V_o \propto \frac{r_B}{t_{NC}} \sqrt{\frac{Z_D Z_{LCP} A_{LCP}}{A_P}} , \tag{22}$$

with a proportionality constant within an order of magnitude of 0.61 MeV×$10^{-22}$s/fm$^2$×$r_B$ as obtained for a linear nuclear potential moving linearly with time (see Eq. (17)).

In the present paper, we are only interested in obtaining evidence that low-energy ternary fission may be due to a coupling of an evaporation-like process with the rapid re-arrangement of the nuclear fluid following scission, and in obtaining an order of magnitude estimate of this re-arrangement time. We are not at this time overly interested in determining the correct complex spatial and temporal dependence of the nuclear potential in the neck region at and following scission. Therefore, given the insensitivity of the functional form of $\Delta V_o$ on the spatial and temporal dependence of the nuclear potential, for simplicity we assume a linear nuclear potential moving linearly with time. Given this simplicity, an analytical expression for the drop in the effective emission barriers (see Eq. (16)) can be obtained without invoking the limit, nuclear force inside the barrier$\to \infty$. As discussed above, depending on the true nature of the nuclear potential around the neck region, at and following scission, the time scales extracted using Eq. (16) could differ significantly from the true neck collapse time.

### II.C Particle emission including excited states

A significant number of the LCP emitted in ternary fission have either no excited states or excited states at such high energies that the probability of non-ground-state emission is negligible. However, a large number of the heavier LCP do have excited states with excitation energies and spins such that calculations including the emission of the excited systems is warranted. Equation (2) represents the ternary fission emission probability for particles in a given state (e.g. the ground state). To calculate the emission probability of a given isotope including the emission of excited states, Eq. (2) must be summed over all excited states, $i$:

$$P_{TF} = C \ A_{LCP} \sum_i (2s_i + 1) \exp\left(\frac{-\Delta E_i}{T_{\text{scis}}}\right) , \tag{23}$$

where the energy costs $\Delta E_i$ are equal to the energy cost for the emission of a particle in its ground state plus the excitation energy of the excited state $i$. In the present work, all known states with an excitation energy up to ~7 MeV are included. With a temperature of ~1.2 MeV, the sum over excited states leads to enhancements relative to purely ground-state emission of more than a factor of two for some systems, with a few enhancement factors greater than 5 for some of the $Z \geq 7$ systems.

### III. Comparison of the model with experiment

The most complete set of measured ternary fission probabilities for a given reaction is the $^{241}$Pu(n$_{th}$,f) data of Köster *et al*. [31]. For this reason, in the present paper we focus on the $^{241}$Pu(n$_{th}$,f) data. The measured equatorial ternary fission probabilities of Köster *et al*. [31] are analyzed via the following procedure. For a given choice of the model parameters, a fitting metric defined by

$$M^2 = \sum_j \{\ln[P_{TF}^{\exp}(Z_j, A_j)] - \ln[P_{TF}(Z_j, A_j)]\}^2 / n, \tag{24}$$

is determined, where $P_{TF}$ are the calculated ternary fission probabilities, $P_{TF}^{\exp}$ are the corresponding measured (experimental) emission probabilities, and $n$ is the number of fitted experimental data points. The exponential of $M$ is a measure of the typical relative difference between the model calculations and the experimental data. For example, if $M$ was ~1 then the average relative discrepancy between model and experiment would be a factor of ~3. The calculated ternary fission probabilities are obtained using Eq. (23) with particle binding energies estimated using Eqs. (3-7,9) as discussed in section II (see table I or model I in table II). The model parameters are then adjusted to minimize the fitting metric defined in Eq. (24). The summation in Eq. (24) is over all LCP with $Z_{LCP} \leq 6$ for which an experimental value exists. As discussed in the introduction, several recent works seem to support a hot ternary-fission process [19-22,24]. These recent results are strong evidence that in low-energy ternary-fission LCP are being produced via a statistical process. Evidence for a temperature of ~1.1 MeV at scission is summarized in table III. Viewed individually, some of the reasons for believing that the scission configuration in ternary fission has a temperature of ~1.1 MeV, are weak. However, viewed as a whole, the evidence summarized in table III strongly implies that scission configurations in low-energy ternary fission have a temperature of ~1.1 MeV, and thus we shall initially assume this temperature when modelling ternary-fission relative emission probabilities via Eq. (23).

TABLE III. Summary of evidence for a temperature of ~1.1 MeV for low-energy ternary fission scission configurations.

| Basis for the inferred temperature | | Inferred Temperature (MeV) |
|---|---|---|
| Mean kinetic energy of fragments | Dynamical calculations like those discussed in ref [24] | 1.1 |
| Isotope thermometry | Ref [19] | 1.10±0.15 |
| $^{10}$Be 3.368 MeV $\gamma$-ray | Ref [21] | 1.0±0.1 |
| ETF yield ratios | Ref [22] | 1.24±0.10 |
| Polar emission ratio | Ref [22] | 1.13±0.24 |

To estimate the standard static-emission barrier heights for equatorial ternary fission we use Eq. (14) with

$$r = r_B = 0.32 \text{ fm} \sqrt{40 - A_{LCP}} + 1.225 \text{ fm} A_{LCP}^{1/3} + 1.2 \text{ fm} . \tag{25}$$

This dependence of the barrier radius on the mass of the LCP is based on the assumptions that the neck radius at scission, for binary fission, is 2 fm, that the neck material is cylindrical and contains ~40 nucleons, and that the nuclear force has a range of 1.2 fm. A range of 1.2 fm was chosen because this is twice the nominal Woods-Saxon diffuseness parameter ($\delta$=0.6 fm). The sensitivity of the model calculations to these assumptions is discussed later. Fig. 6 shows the $^{241}$Pu(n$_{th}$,f) ternary fission emission probability data of Köster *et al.* [31]. The horizontal axis was chosen to spread the data out in such a way that the different elements do not overlap. The solid curves show a model calculation assuming a nuclear temperature at scission $T_{scis}$=1.1 MeV, and using standard static emission barriers estimated using $r_B$ as given in Eq. (25). The model fails to reproduce the experimental ternary fission probabilities. If standard static barriers are used, then no reasonable combination of a nuclear temperature and barrier radius as a function of the mass of the LCP can reproduce the trends seen in the experimental data.

Fig. 7 shows the $^{241}$Pu(n$_{th}$,f) ternary fission emission probability data of Köster *et al.* [31] along with a model calculation with a temperature $T_{scis}$=1.1 MeV, and with the effective barrier heights obtained using the correction associated with the time dependence of the potential, estimated using Eq. (16) with $t_{NC}$=1.6×10$^{-22}$ s and $F_N$=8.5 MeV/fm. These two model parameters are obtained by minimizing the fitting metric defined in Eq. (24). This model calculation is in good agreement with the trends in the experimental data, spanning nearly seven orders of magnitude. The model correctly predicts that tritons are the dominant hydrogen isotope, the peaks at $^4$He, $^{10}$Be, and $^{14}$C, and the magnitude of the pairing effect associated with $Z_{LCP}$. It is important to remember that the free parameters were only adjusted to fit the $Z_{LCP} \le 6$ data, and that the model calculations for $Z_{LCP} > 6$ are an extrapolation of this fit.

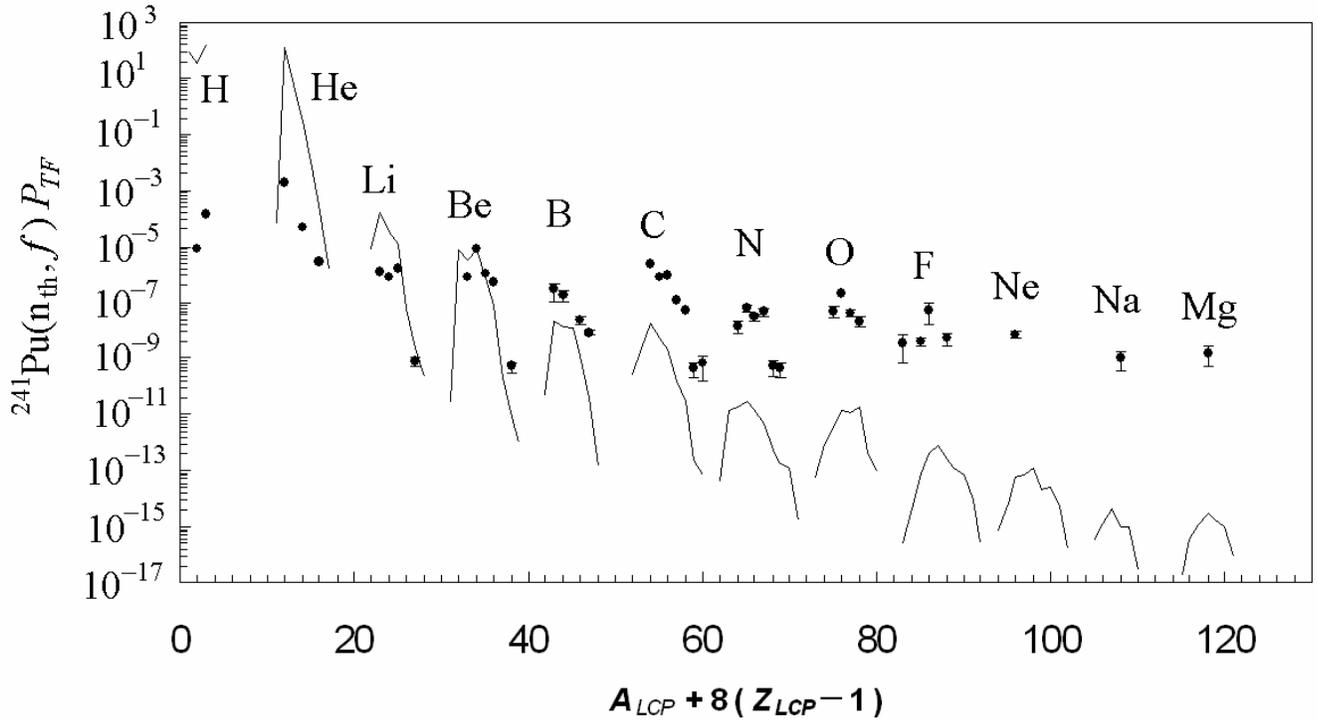

Fig. 6. Measured ternary fission emission probabilities for $^{241}$Pu(n$_{th}$,f) [31]. The solid curves show a model calculation using static emission barriers with $T_{scis}$=1.1 MeV.

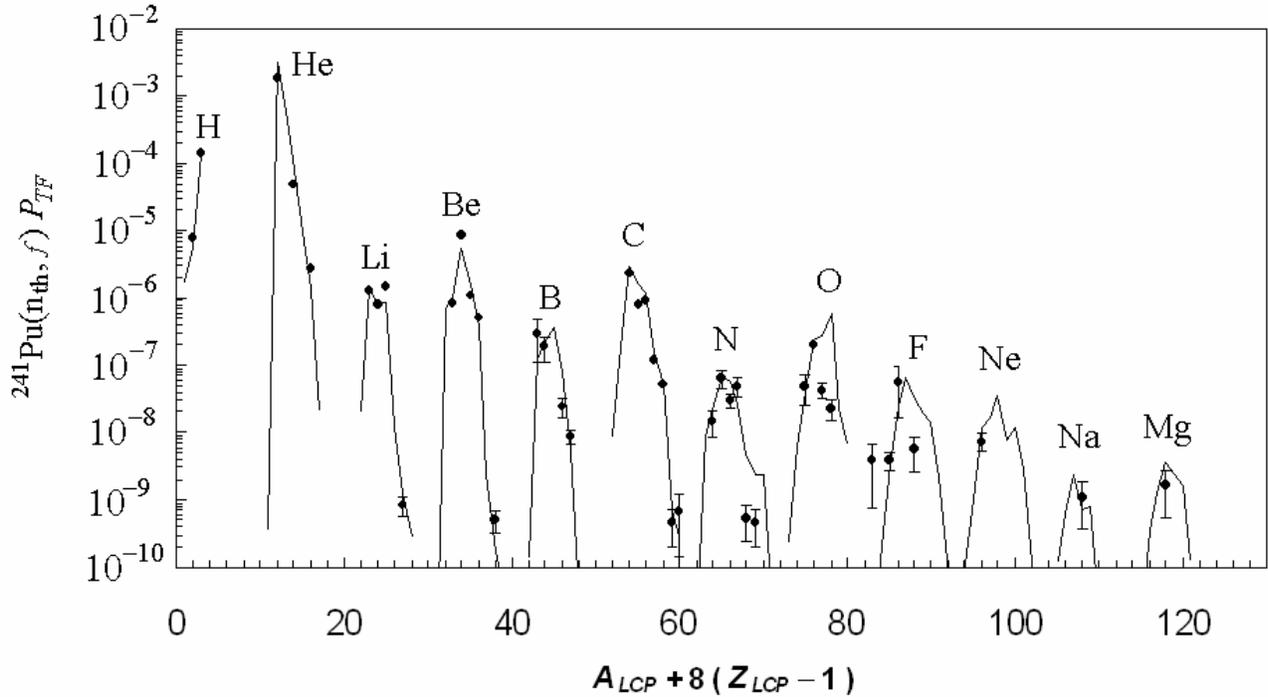

Fig. 7. Measured ternary fission emission probabilities for $^{241}$Pu($n_{th}$,$f$) [31]. The solid curves show a model calculation with $T_{scis}$=1.1 MeV, $t_{NC}$=1.6×10$^{-22}$ s, and $F_N$=8.5 MeV/fm.

A slight improvement in the quality of the agreement between the model and the experimental data can be achieved if the scission temperature, $T_{scis}$, is also treated as an adjustable parameter. Fig. 8 shows the $^{241}$Pu($n_{th}$,$f$) fitting metric, $M^2$, as a function of $T_{scis}$ and $t_{NC}$. Fig. 9 shows the model calculations with a temperature $T_{scis}$=1.19 MeV, $t_{NC}$=1.6×10$^{-22}$ s and $F_N$=8.2 MeV/fm. These three model parameters were obtained by minimizing the fitting metric defined in Eq. (24). The corresponding minimum value for the fitting metric is $M^2$=0.28. Therefore, the typical relative discrepancy between the model calculations and the measured ternary fission probabilities is a little less than a factor of ~2. To reduce the number of free parameters, several quantities have been estimated and fixed. For example, the neck radius at the moment of scission and the range of the nuclear force were set to $r_{neck}$=2 fm, and $N_{range}$=1.2 fm, respectively (see Eq. (25)). If these quantities are now treated as additional free parameters then a very slight improvement in the quality of the fit is obtained with $r_{neck}$=1.7 fm, $N_{range}$=0.6 fm, $T_{scis}$=1.15 MeV, $t_{NC}$=1.1×10$^{-22}$ s and $F_N$=10.0 MeV/fm. The quality of the agreement between the experimental data and the model calculations is also very insensitive to the assumed dependence of the barrier radius on $A_{LCP}$. For example, if $r_B$ is assumed to be 5.2 fm independent of $A_{LCP}$, (and not as given in Eq. (25)) then a good reproduction of the data can be obtained with $T_{scis}$=1.24 MeV, $t_{NC}$=1.6×10$^{-22}$ s and $F_N$=14.4 MeV/fm. Another quantity that has been estimated and fixed is the distance between the mass centers of the nascent fragments at the moment of scission. This quantity affects the calculated particle binding energies via Eqs. (6,7,9) and the static emission barriers via Eq. (14). The distance between mass centers at scission is controlled by the assumed distance between mass centers of the nascent fragments at scission in units of the radius of the corresponding spherical system ($d_{scis}/R_0$=2.6). Setting the range of the nuclear force back to 1.2 fm, and making both $d_{scis}/R_0$ and $r_{neck}$ additional free parameters, results in a very slight improvement in the quality of the fit, with $d_{scis}/R_0$=2.42, $r_{neck}$=2.0 fm, $T_{scis}$=1.15 MeV, $t_{NC}$=1.5×10$^{-22}$ s and $F_N$=8.6 MeV/fm. These results suggest that the ternary fission data are indicating that the neck radius at scission is ~2 fm and that $d_{scis}/R_0$~2.4 (within 10% of the earlier assumed value of 2.6). However, the fitting metric varies weakly with these quantities and, therefore, strong conclusions regarding these quantities can not be drawn at this time. If we assume that the quantities $d_{scis}/R_o$, $r_{neck}$, and $N_{range}$ could lie in the ranges 2.4-2.8, 1.5-2.5 fm, and 0.6-1.8 fm, respectively, then the extracted ranges for the nuclear temperature at scission, the neck collapse time, and the strength of the nuclear force in the neck region are $T_{scis}$=1.2±0.1 MeV, $t_{NC}$=(2±1)×10$^{-22}$ s and $F_N$=8±3 MeV/fm.

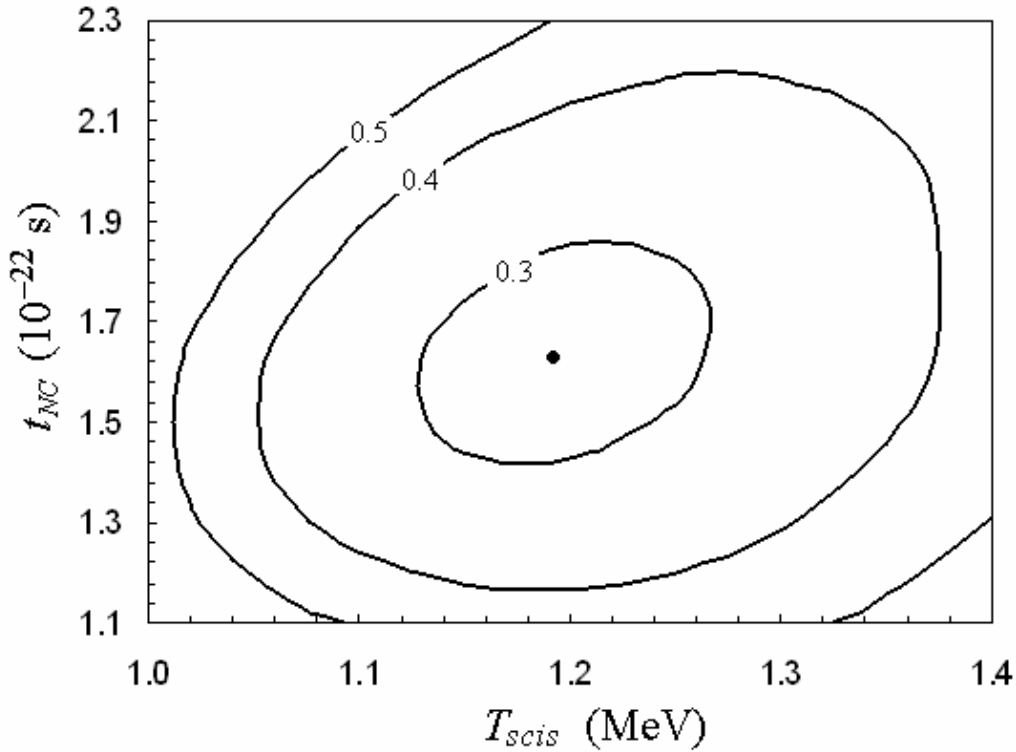

Fig. 8. Contour plot of the $^{241}$Pu(n$_{th}$,$f$) fitting metric, $M^2$, as a function of the temperature at scission ($T_{scis}$) and the neck collapse time ($t_{NC}$). The minimum value of the fitting metric, $M^2$, is 0.28 and its location is shown by the dot.

Fig. 10 shows the measured $^{239}$Pu(n$_{th}$,$f$) ternary fission emission probabilities [32] and the corresponding model calculation without any change to the parameters used to obtain the result shown in Fig. 9. The model calculations reproduce the trends seen in the experimental data from deuterons to oxygen. The model fails to reproduce the proton to deuteron ratio. The reason for this discrepancy is not understood. The ratios of the measured $^{241}$Pu(n$_{th}$,$f$) to $^{239}$Pu(n$_{th}$,$f$) ternary fission probabilities are shown in Fig. 11, along with the corresponding model calculation. The model predicts that for the ternary fission emission of a given element, the lighter isotopes are suppressed and the heavier isotopes are enhanced by an increase in the neutron number of the fissioning system. This behavior is a consequence of the way particle binding energies change as a function of the mass of the fissioning system [22]. Given the accuracy of $^4$He and $^8$He ratios, these data are supportive of the idea that a statistical process plays a central role in low-energy ternary fission. Although the trends in the lithium, beryllium, and carbon data are consistent with the model calculations, new more accurate measurements of the ternary fission emission probabilities for isotopes with $Z_{LCP} \geq 3$ and, in particular, more accurate measurements of the ratio of emission probabilities are needed to either confirm or negate the importance of a statistical process in low-energy ternary fission.

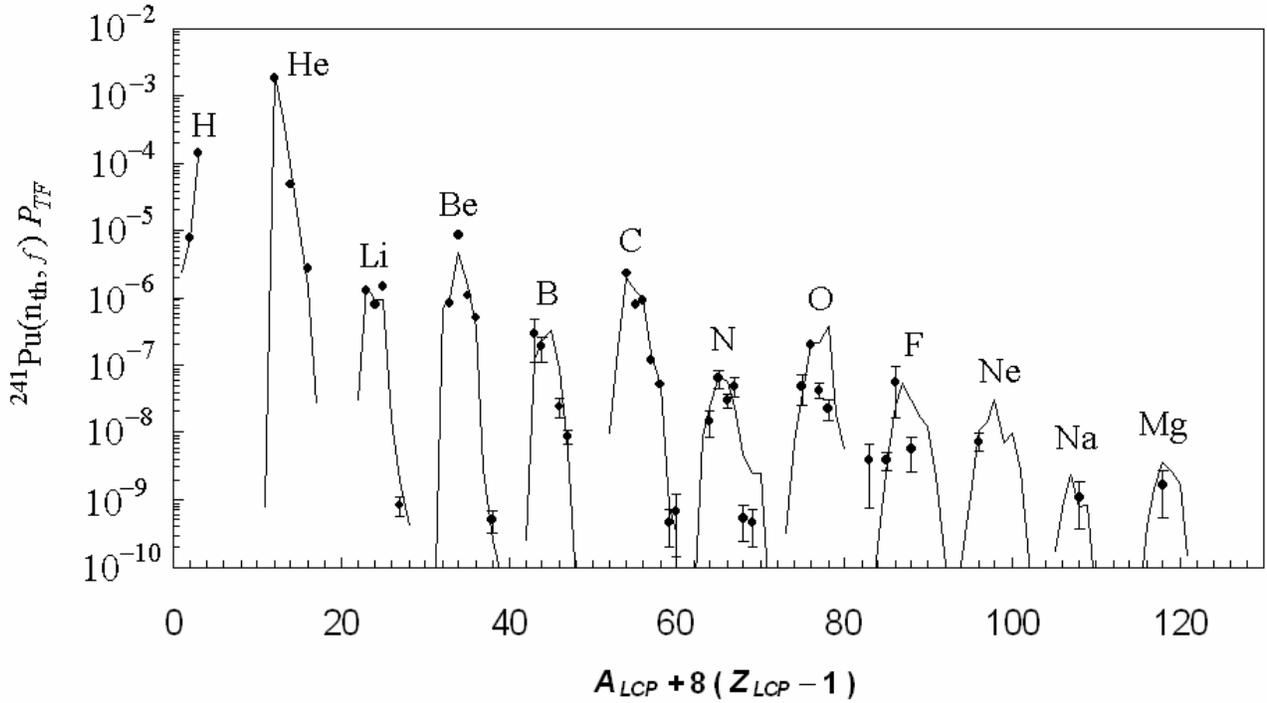

Fig. 9. Measured ternary fission emission probabilities for $^{241}$Pu(n$_{th}$,f) [31]. The solid curves show a model calculation with $T_{scis}$=1.19 MeV, $t_{NC}$=1.6×10$^{-22}$ s, $F_N$=8.2 MeV/fm.

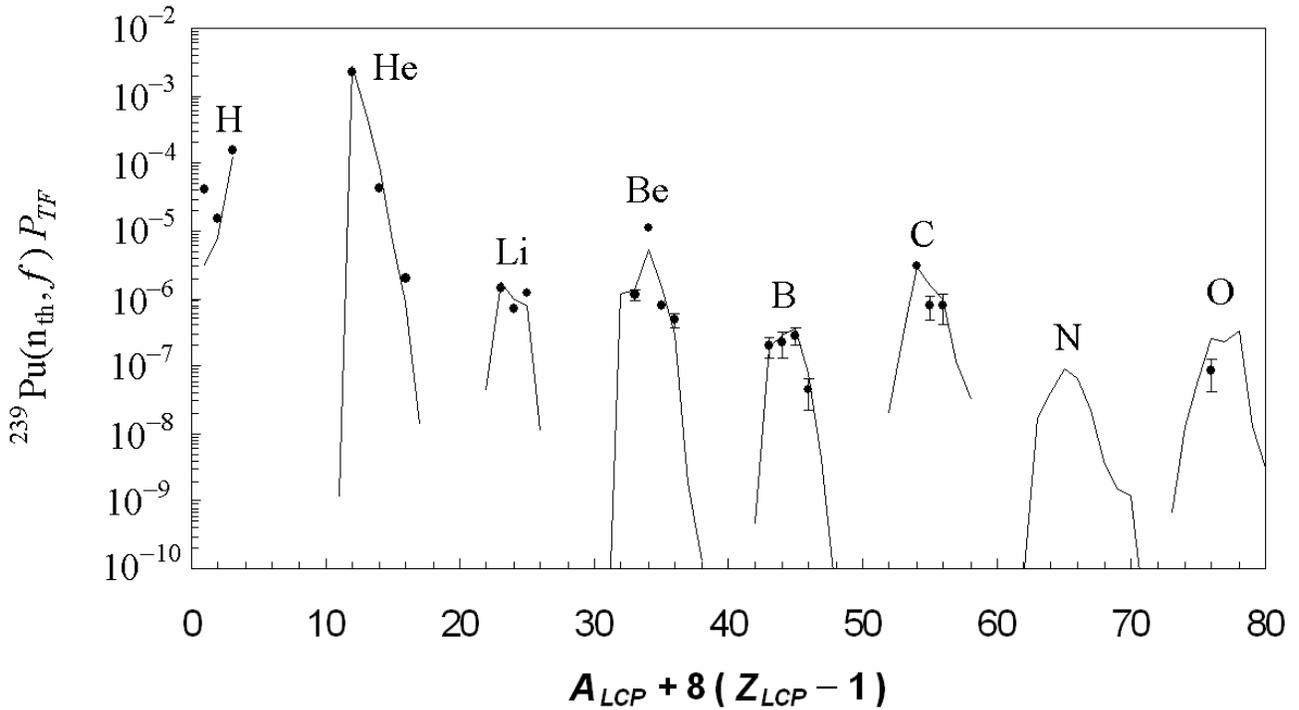

Fig. 10. Measured ternary fission emission probabilities for $^{239}$Pu(n$_{th}$,f) [32]. The solid curves show a model calculation with $T_{scis}$=1.19 MeV, $t_{NC}$=1.6×10$^{-22}$ s, and $F_N$=8.2 MeV/fm.

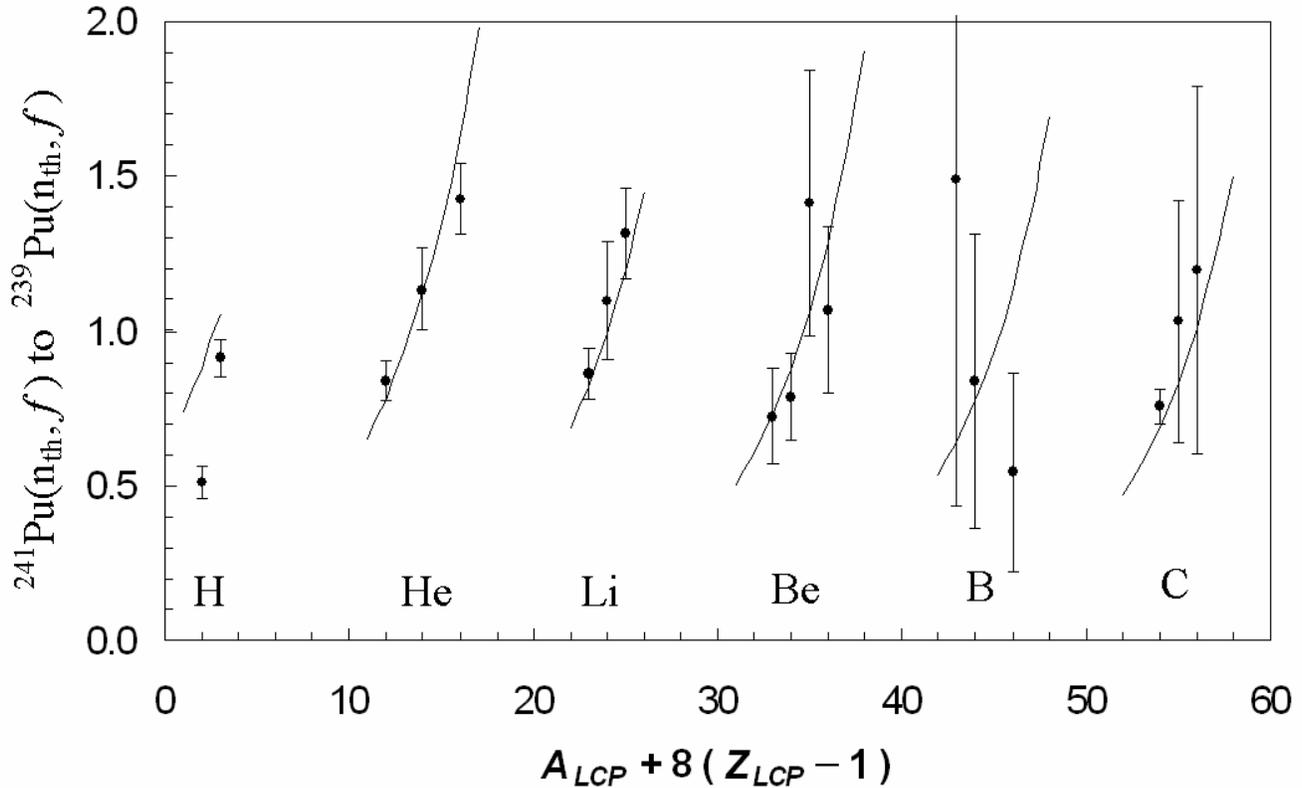

Fig. 11. The ratio of the measured $^{241}Pu(n_{th},f)$ [31] to $^{239}Pu(n_{th},f)$ [32] ternary fission probabilities. The solid curves show the model calculation with $T_{scis}$=1.19 MeV, $t_{NC}$=1.6×10$^{-22}$ s, and $F_N$=8.2 MeV/fm.

## IV. Discussion

There have been previous attempts to use statistical methods to describe ternary fission, for example [33-35]. The key differences with the model presented here are, in the present work the particle binding energies for equatorial ternary fission from the scission configuration are calculated taking into account the change in the distance between mass centers of the nascent parent and daughter fragments produced by the formation of the LCP between the main two nascent fragments; and the effective emission barrier heights are estimated as a function of a dynamical time scale associated with the rapid collapse of the neck material following scission. This evaporation-based model does not suggest that dynamical effects are not important, but rather that, if the time dependence of the nuclear potential felt by LCP in the region between the main nascent fragments is incorporated into a statistical model, then a significantly improved reproduction of the relative ternary fission emission probabilities is obtained relative to a statistical model calculation that contains no dynamical effects.

Simple computational methods have been used here, and thus there is significant room for improvement in the model calculations presented here. The effective barrier heights for ternary fission LCP are defined in terms of the motion of particles that start at rest at scission. It is unclear why this definition of the effective barrier height leads to such a good reproduction of the relative experimental ternary fission emission probabilities. Further theoretical work will be required to determine whether the good agreement is accidental or reflects significant physics. The potential energies of evaporated particles are presently being estimated assuming a linear nuclear potential, which varies with time in a simplistic fashion, and by assuming point particles in the Coulomb field of two spherical fission fragments. It would be more satisfactory to calculate the Coulomb potential energy including the finite size of the LCP in the Coulomb field of a series of deformed shapes that evolve smoothly from the scission configuration to two nearly-spherical fragments following scission. Similarly, the nuclear potential energy could be calculated as a function of time using a folded-Yukawa [36] or folded-exponential potential [37]. This would give a more realistic estimate of the potential energy of LCP in the neck region as a function of LCP $A$, $Z$, location, and time, if the motion of the collapsing neck material could be accurately modeled. Given the simple time dependence of the potentials used here and the fact that changing between different assumed potentials leads to significant changes in inferred time scales, the true neck collapse time might differ significantly from the inferred values. Therefore, the present work can only be used to conclude that the neck collapse time is of the order of $10^{-22}$ s. This neck collapse time scale is reasonable and sits between upper and lower limits for the neck collapse time that can be determined from the time scales for standard collective flow and the Fermi velocity of nucleons. The angular frequency for quadrupole oscillations about the equilibrium position of hot nuclei is ~$10^{+21}$ $s^{-1}$ [28]. The corresponding time scale for a prolate system to collapse into a spherical shape is ~$10^{-21}$ s. The neck collapse involves hydrodynamic instabilities and is likely to be much faster than this time scale. For the neck to collapse, nucleons have to move a distance of ~3 fm (see Fig. 1). The Fermi velocity of nucleons inside standard nuclear matter is ~1 fm per $10^{-23}$ s and therefore it would be difficult to understand how the neck collapse could occur in a time faster than ~$3\times10^{-23}$ s.

When the neck material between the nascent fragments collapses following scission, it is not just the neck material that experiences acceleration. Immediately following scission, each nascent fragment can be thought of as a nearly spherical fragment plus neck material. A small number of particles may be emitted from the outer polar tips of scission configurations while the nearly spherical parts of the nascent fragments accelerate towards the retracting neck stubs. Therefore, the particle emission process described in this paper might also be responsible for polar emission. Given the relative mass of the nearly spherical and neck parts of the nascent fragments, the acceleration of the nearly spherical part of the nascent fragments will be ~1/5 of the acceleration of the neck material. Therefore, the polar emission will be much weaker than the equatorial ternary fission. The equatorial ternary-fission binding-energy shifts given by Eq. (9) favor the ejection of heavier particles, and the effective barrier heights are significantly lower for the heavier isotopes. This is why, in equatorial ternary fission, tritons are the dominant hydrogen isotope, and isotopes heavier than $\alpha$ particles are seen. In the weaker polar ternary-fission process, the particle binding energy shifts will work in the opposite direction [22] and the effective barrier heights will be much closer to the standard static barrier heights. This reasoning is consistent with the observations that, in polar emission, protons are the dominant hydrogen isotope and isotopes heavier than $\alpha$ particles are very rare.

## V. Summary and conclusions

An evaporation-based model of ternary nuclear fission has been applied to thermal-neutron-induced fission of plutonium. This model involves an evaporation process coupled with a rapid collapse of the neck material between the nascent fragments. It is assumed that effective emission-barrier heights can be defined using the motion of LCP that start at rest at scission. Simple one-dimensional calculations are used to describe the motion of LCP in the plane between and perpendicular to the nascent fragments. Following scission, the reach of the nuclear potential around the neck material is assumed to decrease linearly to zero as the neck stubs collapse into the nascent fragments. Calculated relative ternary fission probabilities for the emission of a wide range of isotopes are in agreement with the trends in the experimental results spanning seven orders of magnitude, if the temperature of scission configurations and the neck collapse time are assumed to be ~1.2 MeV and ~$10^{-22}$ s, respectively. A similar particle ejecting mechanism might also occur at the tips of the scission configuration, producing LCP in the direction of the main fragments.

Based on the measured ternary-fission relative emission probabilities that are reproduced by these model calculations, and that the required nuclear temperature at scission is consistent with independent estimates obtained by other means, it is concluded that low-energy ternary fission may not be inconsistent with an evaporation-like process coupled to the rapid re-arrangement of the nuclear fluid following scission.

## Appendix A: Analytical expressions for the effective emission barriers

In particle evaporation from a static spherical system, the locations (and thus heights) of the emission barriers can be determined by considering the trajectory of test particles starting at rest. If these test particles are placed beyond the barrier location ($r > r_B$) then the particles travel out to large radii. If test particles are started inside the barrier location ($r < r_B$) then the particles travel to $r=0$. Here we assume the same definition can be used to define the effective barrier location, $r_{EB}$, for equatorial ternary fission. Consider a LCP starting at rest inside the static barrier location of the scission configuration, on the plane perpendicular to and symmetrically between two symmetric nascent fragments at the moment of scission, a distance $r_i = r_B - \Delta r$ from the symmetry axis. Given the symmetry of the starting location, the motion of the particle can be solved in one dimension. Many of the following assumptions are very simplistic and made in the spirit of finding an analytical expression for the effective emission barrier height. We shall assume that the nuclear force is zero beyond the location of the static barrier, $r_B$, and a constant $F_N$ inside the location of the static barrier. Inside the static barrier location, it is assumed that the nuclear force dominates and that Coulomb repulsion is, in comparison, negligible. Fig. A1 is a schematic representation of the assumed potential of a LCP in the region of the neck material at the time of scission. The small solid circle represents the starting location of a test particle.

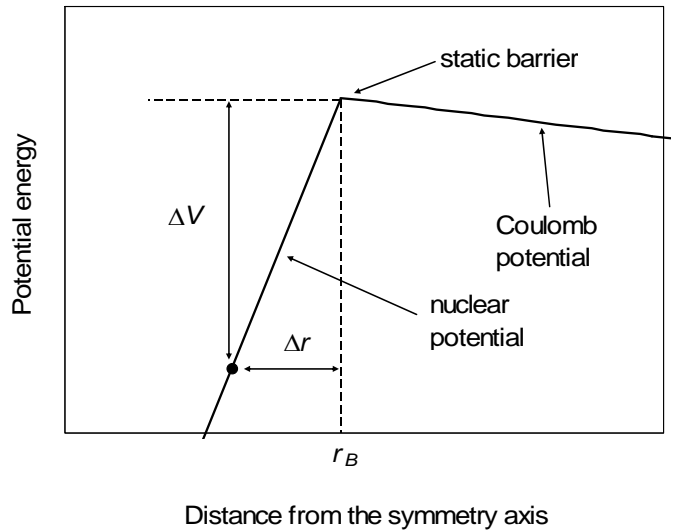

Fig. A1. A schematic representation of the assumed potential of a LCP in the region of the neck at the time of scission. The small solid circle represents the starting location of a test particle.

After scission the reach of the nuclear force is assumed to vary linearly from $r_B$ at the time of scission to zero. This collapsing of the reach of the nuclear force is assumed to occur over a time period $t_{NC}$ (the neck collapse time). While the test particle remains inside the static barrier, its distance from the symmetry axis as a function of time is given by

$$r_p(t) = r_B - \Delta r - \frac{1}{2}\frac{F_N}{m}\cdot t^2, \tag{A-1}$$

where $m$ is the mass of the test particle. The location of the static barrier as a function of time is assumed to be

$$r_B(t) = r_B - \frac{r_B}{t_{NC}}\cdot t. \tag{A-2}$$

By setting $r_p(t) = r_B(t)$ the time, $t_C$, it takes the static barrier to overtake (catch) the particle can be determined.

$$t_C = \frac{m\,r_B}{F_N\,t_{NC}} - \sqrt{\left(\frac{m\,r_B}{F_N\,t_{NC}}\right)^2 - \frac{2m\,\Delta r}{F_N}} \tag{A-3}$$

The velocity of the test particle as the static barrier overtakes it is then given by

$$v_p(t_C) = \frac{r_B}{t_{NC}} - \sqrt{\left(\frac{r_B}{t_{NC}}\right)^2 - \frac{2\,F_N\,\Delta r}{m}}. \tag{A-4}$$

The kinetic energy of the test particle as the static barrier overtakes it is then given by

$$E_p(t_C) = E_B\left(1 - \sqrt{1 - \frac{\Delta V}{E_B}}\right)^2, \tag{A-5}$$

where $E_B$ is the kinetic energy of the LCP if it was moving with the velocity of the barrier and is given by

$$E_B = \frac{m}{2}\left(\frac{r_B}{t_{NC}}\right)^2. \tag{A-6}$$

The Coulomb potential energy at the time of scission and near the symmetry axis in the plane between the two main fragments is assumed to be dominated by the main fragments and can be approximated as

$$C_P(r) \sim 0.72\,(\text{MeV fm})\frac{Z_{LCP}\,Z_D}{r_{\text{scis}}^3}\left(2\,r_{\text{scis}}^2 - r^2\right), \tag{A-7}$$

where $r_{\text{scis}}$ is half the distance between mass centers of the nascent fragments at scission, and $Z_{LCP}$ and $Z_D$ are the atomic numbers of the LCP and daughter system, respectively. If the half distance between the main fragments at scission is assumed to be $r_{\text{scis}} = 1.3 \times 1.225\,\text{fm} \times A_P^{1/3}$, where $A_P$ is the mass number of the parent system, and if the neck collapse is rapid enough that the center of mass of the main fragments does not significantly move during the neck collapse, then the difference in the Coulomb potential energy of the test particle at its starting location $r_i = r_B - \Delta V/F_N$ and at $r=0$, is given by

$$\Delta C_P = 0.178\,(\frac{\text{MeV}}{\text{fm}^2})\frac{Z_{LCP}\,Z_D}{A_P}\left(r_B - \frac{\Delta V}{F_N}\right)^2. \tag{A-8}$$

We assume that the effective emission barrier is defined by the starting location of the test particle that just makes it to the symmetry axis at $r=0$. Starting locations with $r_i \leq r_{EB}$ reach $r=0$, while starting locations with $r_i > r_{EB}$ approach $r=\infty$ as the time approaches infinity. The value of $\Delta V$ corresponding to the effective barrier location can be determined by setting the kinetic energy gained by the action of the nuclear force given in Eq. (A-5) to the gain in Coulomb potential energy associated with the passage to $r=0$ given in Eq. (A-8). The resulting expression is

$$0 = \left(\frac{r_B}{X} - \frac{\Delta V}{F_N X}\right)^2 - 2\left(\frac{r_B}{X} - \frac{\Delta V}{F_N X}\right) + \frac{\Delta V}{E_B}, \tag{A-9}$$

where

$$X = 2.37 \frac{\text{fm}}{\sqrt{\text{MeV}}} \sqrt{\frac{E_B A_P}{Z_{LCP} Z_D}}. \tag{A-10}$$

Algebraic manipulation of Eq. (A-9) leads to the following quadratic expression for $\Delta V$,

$$0 = \frac{1}{2} \Delta V^2 + \left( F_N X + \frac{F_N^2 X^2}{2 E_B} - F_N r_B \right) \Delta V + \frac{1}{2} \left( F_N^2 r_B^2 - 2 F_N^2 X r_B \right), \tag{A-11}$$

for which the solution is

$$\Delta V = \frac{2 F_N r_B E_B}{\Delta V_o} \left( \frac{\Delta V_o}{2 E_B} - \left( 1 + \frac{F_N r_B}{\Delta V_o} \right) + \sqrt{\left( 1 + \frac{F_N r_B}{\Delta V_o} \right)^2 - \frac{F_N r_B}{E_B}} \right), \tag{A-12}$$

where $\Delta V_o$ is

$$\Delta V_o = 0.61 \frac{\text{MeV} \, 10^{-22} \, \text{s}}{\text{fm}^2} \frac{r_B^2}{t_{NC}} \sqrt{\frac{Z_D Z_{LCP} A_{LCP}}{A_P}}. \tag{A-13}$$

In the limit of a rapid neck collapse, the first term in (A-9) becomes negligible and $\Delta V$ can be approximated by

$$\Delta V = \frac{1}{\frac{1}{F_N r_B} + \frac{1}{\Delta V_o}}, \tag{A-14}$$

and therefore $\Delta V_o$ gives the decrease in the barrier heights in the limits $t_{NC} \to 0$ and $F_N \to \infty$. The analytical expression (A-12) for the drop in the effective barrier heights relative to the static barriers at scission, associated with a finite neck collapse time, was obtained relatively easily because the assumed nuclear force on the LCP is only dependent on the location of the LCP relative to $r_B(t)$. For more realistic spatial dependencies of the nuclear potential, the situation is more complex. However, in the limit as the neck collapse time approaches zero, and if the nuclear potential drops very sharply inside the static barrier location, then analytical expressions can be obtained with complex spatial and temporal dependencies of the nuclear potential. For example, if the nuclear potential around the neck, at and following scission, is assumed to be of the form of an exponential shifting linearly in time:

$$V_N(r,t) = -V_o \, \exp\left( \frac{r_o \left( 1 - t / t_{NC} \right) - r}{\delta} \right), \tag{A-15}$$

then in the limit $t_{NC} \to 0$, the LCP does not move during the neck collapse and the impulse given to the LCP, starting at the effective barrier location, by the nuclear force is given by

$$I = \int \frac{V_N(r_i, 0)}{\delta} \exp\left( \frac{-r_o \, t}{\delta \, t_{NC}} \right) dt \sim \frac{\Delta V_o \, t_{NC}}{r_o}, \tag{A-16}$$

and thus the kinetic energy gained through the action of the nuclear force will be

$$E_p \sim \frac{\Delta V_o^2 \, t_{NC}^2}{2 \, m \, r_o^2}. \tag{A-17}$$

If the nuclear force decreases rapidly inside the static barrier radius, $r_B$, then $\Delta r$ will be small (see Fig. A-1) and the difference in the Coulomb potential energy of the test particle at its starting location $r_i$ and at $r=0$, is given by

$$\Delta C_P \sim 0.178\,(\frac{\text{MeV}}{\text{fm}^2})\,\frac{Z_{LCP}\,Z_D}{A_P}\,r_B^2, \qquad (A\text{-}18)$$

and therefore

$$\Delta V_o \sim 0.61\,\frac{\text{MeV}\,10^{-22}\text{s}}{\text{fm}^2}\,\frac{r_o\,r_B}{t_{NC}}\,\sqrt{\frac{Z_D Z_{LCP} A_{LCP}}{A_P}}\,. \qquad (A\text{-}19)$$

In a similar fashion, it can be shown that if the nuclear potential around the neck at and following scission is assumed to be of the form of an exponential weakening linearly with time:

$$V_{\text{N}}(r,t) = -V_o\left(\frac{t_{NC}-t}{t_{NC}}\right)\exp\left(\frac{r_o-r}{\delta}\right), \quad (t \le t_{NC}), \qquad (A\text{-}20)$$

then in the limits as $t_{NC} \to 0$ and $V_o \to \infty$ the drop in the effective barrier heights relative to the static barriers at scission is given by

$$\Delta V_o \sim 1.22\,\frac{\text{MeV}\,10^{-22}\text{s}}{\text{fm}^2}\,\frac{\delta\,r_B}{t_{NC}}\,\sqrt{\frac{Z_D Z_{LCP} A_{LCP}}{A_P}}\,. \qquad (A\text{-}21)$$

## Acknowledgment

I wish to thank A. J. Sierk for the lengthy discussions we had, and for his valuable input during the development of the models and concepts discussed herein.